\documentclass{pramana}

\usepackage{graphicx,amsmath,bm}

\usepackage{color,soul}

\begin{document}

\title{Nonstandard Finite Difference Time Domain (NSFDTD) Method for Solving the Schr\"odinger Equation}

\author{I Wayan Sudiarta}

\affilOne{Physics Study Program, Faculty of Mathematics and Natural Sciences, University of Mataram, Mataram, NTB, Indonesia}

\twocolumn[{

\maketitle

\corres{wayan.sudiarta@unram.ac.id}
\msinfo{3 January 2018}{3 January 2018}{3 January 2018}

\begin{abstract}
In this paper, an improvement of the finite difference time domain (FDTD) method using a non-standard finite difference scheme for solving the Schr\"odinger equation is presented. The standard numerical scheme for a second derivative in the spatial domain is replaced by a non-standard numerical scheme. In order to apply the non-standard FDTD (NSFDTD), first estimates of eigen-energies of a system are needed and computed by the standard FDTD method. These first eigen-energies are then used by the NSFDTD method to obtain improved eigen-energies. The NSFDTD method can be performed iteratively using the resulting eigen-energies to obtain more accurate results. In this paper, the NSFDTD method is validated using infinite square well, harmonic oscillator and Morse potentials. It is shown that significant improvements are found when using the NSFDTD method. 
\end{abstract}

\keywords{Time-dependent Schr\"odinger equations, Nonstandard Finite difference method, FDTD method,  Eigenvalues and eigenfunctions}

\pacs{02.70.-c; 02.60.Jh; 07.05.Tp; 03.65.Ge}

}]

\doinum{??/??}
\artcitid{\#\#\#\#}
\volnum{?}
\year{2018}
\pgrange{1--7}
\setcounter{page}{1}
\lp{7}

\section{Introduction}

The finite difference time domain (FDTD) method has been used extensively to simulate various electromagnetic wave radiations and interactions \cite{taflove, taflove2013}. Because of its usefulness, easy in programming, good accuracy, and flexibility, the FDTD method has been applied to other fields such as sound wave propagations \cite{botteldooren1995} and simulations of quantum systems  \cite{sullivan2001, sudiarta2007}. 

To simulate a quantum system in real time using the FDTD method, one can use split method where the time-dependent Schr\"odinger equation (TDSE) is splitted into two equations for real and imaginary parts \cite{visscher1991} or one can also directly use complex computation without splitting the TDSE \cite{askar1978}. Eigen values and eigen functions can not be directly obtained from the real-time FDTD method. It requires a Fourier transformation procedure to obtain eigenvalues and an additional FDTD simulation to compute the eigen functions. Beside that for infinite spatial domain, the FDTD method requires also an absorbing boundary condition (ABC) to remove unwanted reflection from the computational boundary \cite{sullivan2002, sullivan2005}. 

Sudiarta and Geldart \cite{sudiarta2007} showed that the Fourier transformation procedure and the ABC are not needed when the TDSE in imaginary time is used to compute eigen energies and eigen functions. The TDSE in real time is transformed into the TDSE in imaginary time (or a diffusion equation) by replacing the real time variable with an imaginary time variable. The diffusion equation is then used to simulate a decaying of a wave function to the ground state wave function. The ground state energy and the ground state wave function can be computed directly in single simulation run. Beside that another advantage of the FDTD method in imaginary time is that it uses only real values of the wave function therefore it requires less computer memory.

The FDTD method in imaginary time (called simply the FDTD method for the rest of this paper) has been applied for various quantum systems. Sudiarta and Geldart \cite{sudiarta2008} have used the FDTD method for an electron in a quantum dot and in a magnetic field. Dumitru et al. \cite{dumitru2009quarkonium} have studied quarkornium states in QCD plasma. Strickland and Yager-Elorriaga \cite{strickland2010parallel} have used the FDTD method with parallel algorithm for a cumputer cluster to simulate a large quantum system. Margotta et al \cite{margotta2011quarkonium} have used the FDTD method to study quarkonium binding energies using a realistic complex-valued potential. The FDTD method has also been used by Alford and Strickland \cite{alford2013charmonia} to study quarkonium states in an external magnetic field. Sudiarta and Geldart \cite{sudiarta2009} have applied the FDTD method to compute single particle density matrices. Sudiarta and Angraini (2016) have also applied the FDTD method with the super symmetry technique to obtain ground and excited states of a particle in one dimensional potentials. 
 
For a quite small spatial grid spacing, the FDTD method has been found to give accurate results for one to three dimensional quantum systems  \cite{sudiarta2007}. For larger systems the FDTD method with small spacing is not possible due to limited computer memory. One can overcome this problem by using a cluster of computers and doing computation in parallel which can be easily done for the FDTD method as presented by Strickland and Yager-Elorriaga \cite{strickland2010parallel}. 

Another method to overcome this limitation is to increase its accuracy by using a nonstandard finite difference (NSFD) scheme that was found to be successfully applied for various differential equations \cite{mickens1992, mickens2000, aydin2017}. The NSFD scheme for the Schr\"odinger equation was applied by Mickens and Ramadhani \cite{mickens1992}. It is shown by Mickens and Ramadhani \cite{mickens1992} and Chen et al. \cite{chen1993} that the nonstandard finite difference scheme is performing better than the standard method and it is also found to be exact for a constant potential. Application of the FDTD method with the NSFD scheme has not been done previously. In this paper, the NSFD scheme is used to modify the standard FDTD method such that the nonstandard FDTD (or NSFDTD) method can be used for large spatial grid spacing.

The NSFD scheme is generally applied for a system with one spatial dimension. Extension to two or more spatial dimensions is currently in progress and it will be published in our future publication. One possible extension is given by Cole \cite{cole1997,cole2002} for electromagnetic waves. In this paper only the theory and numerical method of the NSFDTD method for one spatial dimension are presented.  

This paper is organized as follows: the theory of the NSFDTD method is introduced in the next section, numerical results of energies and wavefunctions for three potential wells are given in section three and the conclusion is given in section four.

\section{Theory}

The time-dependent Schr\"odinger equation (TDSE) in imaginary time for a system of one particle in a one dimensional potensial $V(x)$ using atomic units ($\hbar = m = 1$) is given by \cite{sudiarta2007}
\begin{equation}
\frac{\partial \psi(x,t)}{\partial t} = - \hat{H}\psi(x,t) = -\left[-\frac{1}{2}\frac{\partial^2 }{\partial x^2} + V(x)\right]\psi(x,t)
\label{eqn-tdse}
\end{equation}

The corresponding time-independent Schr\"odinger equation (TISE), which is an eigen value equation, is given by
\begin{equation}
\frac{\partial^2 \psi(x,t)}{\partial x^2} + W(x)\psi(x,t) = 0
\label{eqn-tise}
\end{equation}
where $W(x) = 2[E-V(x)]$. It is noted that $E$ is the eigen energy.

A solution of the TISE can be obtained iteratively by using Eq.~(\ref{eqn-tdse}). Using an arbitrary initial wavefunction $\psi(x,0)$ and a suitable discretization scheme, a simulation with the discretized Eq.~(\ref{eqn-tdse}) can be done to evolve the initial wavefunction. After a long enough simulation time, the wavefunction converges to the ground state of the system. Excited states of the system can be obtained in turn by evolving the initial wavefunction with the component of lower energy wavefunctions removed from the wavefunction \cite{sudiarta2007}. For computational purpose, the computational domain is limited and the boundary of computation domain is terminated by a Dirichlet boundary condition, $\psi(boundary) = 0$. This boundary condition is equivalent to the infinite potential wall. After the wavefunction is obtained, the corresponding energy is computed with 

\begin{equation}
E  = \frac{\int \psi^*(x)\hat{H} \psi(x) dx}{\int |\psi(x)|^2 dx}
\end{equation}

In the FDTD method \cite{sudiarta2007}, the standard discritization for the second derivative of Eq.~(\ref{eqn-tdse}) is the central finite difference scheme \cite{chapra} in the form of  

\begin{equation}
\frac{\partial^2 \psi(x)}{\partial x^2} \approx \frac{\psi(i+1) - 2\psi(i) + \psi(i-1)}{(\Delta x)^2}
\label{eqn-std}
\end{equation}
where we have used a notation $\psi(i) \equiv \psi(i\Delta x)$ and $\Delta x$ is the spatial grid spacing. 

For certain applications, the NSFD scheme developed by Mickens \cite{mickens1999, mickens2002} has been shown to give exact numerical results. A nonstandard scheme for the Schr\"odinger equation given by Mickens and Ramadhani \cite{mickens1992} can be found by using a Taylor series for finite difference of the second derivative as the following:  
\begin{eqnarray}
\psi(i+1) - 2\psi(i) + \psi(i-1) = \hspace{5em}\nonumber \\ 
2\left[\frac{(\Delta x)^2}{2!}\psi'' + \frac{(\Delta x)^4}{4!}\psi^{(4)} + \frac{(\Delta x)^6}{6!}\psi^{(6)} + \ldots \right]\label{eqn-taylorfd}
\end{eqnarray}
where $\psi'' = \partial^2 \psi/\partial x^2$, $\psi^{(4)} = \partial^4 \psi/\partial x^4$, and $\psi^{(6)} = \partial^6 \psi/\partial x^6$.  
Equation (\ref{eqn-tise}) with a constant potential, $W(x) = W$, can be used to simplify Eq. (\ref{eqn-taylorfd}). It can be shown that
\begin{eqnarray}
\psi(i+1) - 2\psi(i) + \psi(i-1) = \hspace{8em}\nonumber\\
2\frac{\psi''}{W}\left[\frac{(\Delta x \sqrt{W})^2}{2!} - \frac{(\Delta x\sqrt{W})^4}{4!} + \frac{(\Delta x\sqrt{W})^6}{6!} - \ldots\right]
\label{eqn-taylorfd2}
\end{eqnarray}
Equation (\ref{eqn-taylorfd2}) can be simplified further using Maclaurin series of $2\sin^2(\theta/2)$, we obtain  
\begin{equation}
\psi(i+1) - 2\psi(i) + \psi(i-1) 
= \psi''\left[\frac{4}{W}\sin^2\left(\sqrt{W}\Delta x /2\right)\right]
\label{eqn-nsfd}
\end{equation}

Therefore, the second derivative can be accurately approximated by  
\begin{equation}
\frac{\partial^2 \psi(x)}{\partial x^2} \approx \frac{\psi(i+1) - 2\psi(i) + \psi(i-1)}{g(i)}
\label{eqn-nonstd}
\end{equation}
where the denominator function $g(i) \equiv g(i\Delta x) = g(x)$ is given by
\begin{equation}
g(x) = \left\{\begin{array}{ll}
(\Delta x)^2 & \text{for} \ W(x) = 0 \\
\frac{4}{W(x)}\sin^2\left(\sqrt{W(x)}\Delta x/2\right) & \text{for}\ W(x) > 0 \\
\frac{4}{-W(x)}\sinh^2\left(\sqrt{-W(x)}\Delta x/2\right) & \text{for}\ W(x) < 0 
\end{array} \right.
\label{eqn-gx}
\end{equation}.
$g(x)$ can be rewritten in a simpler form using a Taylor series as
\begin{equation}
g(x) \approx (\Delta x)^2 - \frac{W(x)}{12}(\Delta x)^4 + \frac{[W(x)]^2}{360}(\Delta x)^6
\label{eqn-gxtaylor}
\end{equation}  

It is noted in Eq.~(\ref{eqn-gxtaylor}) that the NSFD scheme reduces to the standard scheme when the value $W(x)$ is closed to zero or the spatial interval $\Delta x$ is small. The NSFD scheme of Eq.~(\ref{eqn-nonstd}) is found to give exact results for Eq.~(\ref{eqn-tise}) when $W(x)$ has no spatial dependence or a constant as indicated by Eq. (\ref{eqn-nsfd}). This is in agreement with Mickens and Ramadani \cite{mickens1992} and Chen et al. \cite{chen1993}. When $W(x)$ is not a constant, the nonstandard scheme has accuracy of the order $(\Delta x)^2$ and known to perform better than other methods such as the standard finite difference method and the Numerov method for large values of $\Delta x$ \cite{chen1993}. 

In order to apply the NSFDTD scheme, initial values of eigen energies $E$ are required as input parameters. The first estimates of energies can be obtained by using the standard FDTD method or by using $E = 0$. The values of eigen energies are then improved by using the NSFDTD method. Further improvement can be done by reapplying the NSFDTD method. This proses can be iterated until converged energies are obtained.   

\section{Numerical Methods}

An explicit iterative procedure for the NSFDTD method is following Sudiarta and Geldart \cite{sudiarta2007}. The temporal derivative in Eq.~(\ref{eqn-tdse}) is approximated by the forward difference equation and the spatial second derivative is approximated using Eq.~(\ref{eqn-nonstd}). For better numerical stability and accuracy, $\psi(x,t)$ in Eq.~(\ref{eqn-tdse}) is equal to an average value $(\psi^{n+1}(i) + \psi^n(i))/2$. After some manipulations, it is found that an explicit equation is given by   

\begin{equation}
\psi^{n+1}(i) = \alpha(i) \psi^n(i) +  \frac{\beta(i)\Delta t}{2g(i)}[\psi^n(i+1) - 2\psi^n(i) + \psi^n(i-1)]
\label{eqn-iter}
\end{equation} 
\noindent where $\alpha(i) = [1 - \Delta t V(i)]/[1+\Delta t V(i)]$ and $\beta(i) = 1/[1+\Delta t V(i)]$.

Using initial random values for an initial wavefunction $\psi(i)$, we then iterate with Eq.~(\ref{eqn-iter}) and after enough number of iterations, the wavefunction converges to the groundstate wavefunction. Excited states are obtained by redoing the iteration with lower states removed from the wavefunction (see Sudiarta and Geldart \cite{sudiarta2007} for detail).

The energy is computed numerically by
\begin{eqnarray}
E &&= \frac{1}{\sum_i \psi(i)^2} \times \nonumber \\
&& \sum_i\left\{ V(i)\psi(i)^2 - \frac{\psi(i)[\psi(i+1) - 2\psi(i) + \psi(i-1)]}{g(i)}\right\} \nonumber \\
\end{eqnarray}.

As previously mentioned  that in order to use the NSFDTD method, initial eigen energies are needed and in this paper we obtain them by performing iteration with $g(i) = (\Delta x)^2$ or setting $W(x) =0$ in Eq. (\ref{eqn-gx}). The resulting energies and wavefunctions for this case are the same as the numerical results found using the standard FDTD method. Alternatively we can also use $E_n = 0$ as the initial eigen energies and it is found that same final results of the NSFDTD method are obtained regardless the values of the initial eigen energies. In this paper, only numerical results using $W(x) =0$ are given.

\section{Results and Discussion}

\subsection{Infinite Square Well Potential}

To demonstrate significant improvements when using the NSFDTD in particular for a constant potential $V(x) = V_0$ or simply $V(x) = 0$, we first consider a free particle in an infinite square well given by $V(x) = 0$ for $0<x<1$ and $V(x) = \infty$ for $x<0$ and $x>1$. The eigen energies are known to be $E_n = n^2\pi^2/2$ and the eigen states are given by $\psi_n(x) = \sqrt{2}\sin(n\pi x)$. The computational domain is terminated using $\psi(x) = 0 $ at $x=0$ and $x=1$ which corresponds to an infinite potential wall.

\begin{figure}[p]
\begin{center}
\includegraphics[scale=1.0]{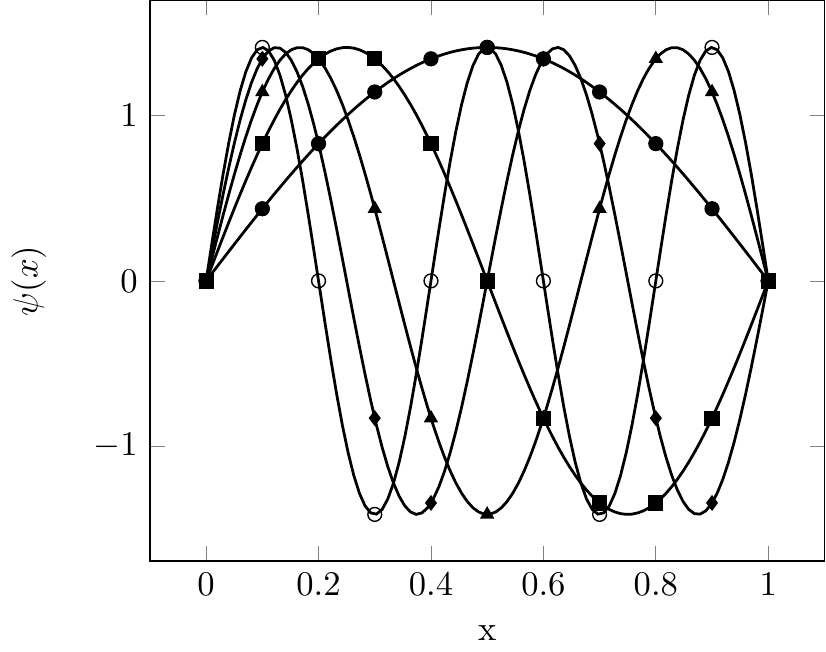}
\end{center}
\caption{Comparison of ground and excited state wavefunctions (symbols: for state n = 1 ($\bullet$), n = 2 ( $\blacksquare$), n = 3  ($\blacktriangle$), n = 4 ($\blacklozenge$), n = 5 ($\circ$)) with analytical wavefunctions (lines) for a particle in an infinite square well.}
\label{fig-square}
\end{figure}

The numerical parameters used for computation are $\Delta x = 0.1$ and $\Delta t = (\Delta x)^2/10$.  The numerical results for eigen energies and state functions are shown in Table \ref{tbl-square} and Fig.~\ref{fig-square}. 

It is noted that after the first iteration the NSFDTD results have shown better accuracy than the standard FDTD method. Further reapplication of the NSFDTD method produces even better eigen energies. After about 10 iteration, the final eigen energies converges to the exact values. These results confirm that the NSFDTD method is exact method when the potential is constant. Furthermore, the resulting wavefunctions are in good agreement with the analytical wavefunctions and the same as the standard FDTD results. This is a consequence of the fact that the values of $g(i)$ are constants. Discritization of the TISE using the standard and nonstandard schemes yields difference equations that differ only in an energy scaling factor.      
  
\begin{table*}[p]
\caption{\label{tbl-square} Numerical eigen-energies for a particle in an infinite square well potential computed by the FDTD method (standard scheme),  the NSFDTD method after first (1st), second (2nd), third (3rd) and final iteration are compared with exact values. The NSFDTD parameters used are $\Delta x = 0.1$ dan the length of the square well or the computational domain is 1. The correct numbers are in bold.} 
\begin{center}
\begin{tabular}{@{}rrrrrrr}
\hline
n	&	FDTD	&	\multicolumn{4}{c}{NSFDTD}&	Exact	\\
	&		&	1st	&	2nd	&	3rd	&	final	&		\\
\hline
1	&	\textbf{4}.894348	&	\textbf{4.934}469	&	\textbf{4.934}799	&	\textbf{4.934802}	&	\textbf{4.934802}	&	4.934802	\\
2	&	\textbf{19}.098301	&	\textbf{19.7}17997	&	\textbf{19.73}8506	&	\textbf{19.739}186	&	\textbf{19.739209}	&	19.739209	\\
3	&	\textbf{4}1.221475	&	\text{44}.174164	&	\textbf{44}.395261	&	\textbf{44.41}1870	&	\textbf{44.413220}	&	44.413220	\\
4	&	69.098301	&	\textbf{7}7.637465	&	\textbf{78}.778707	&	\textbf{78.9}32758	&	\textbf{78.956835}	&	78.956835	\\
5	&	\textbf{1}00.000000	&	\textbf{1}18.475509	&	\textbf{12}2.325053	&	\textbf{123}.146042	&	\textbf{123.37005}0	&	123.370055	\\
\hline
\end{tabular}
\end{center}
\end{table*}

\subsection{Harmonic Oscillator Potential}

\begin{figure}[p]
\begin{center}
\includegraphics[scale=1.0]{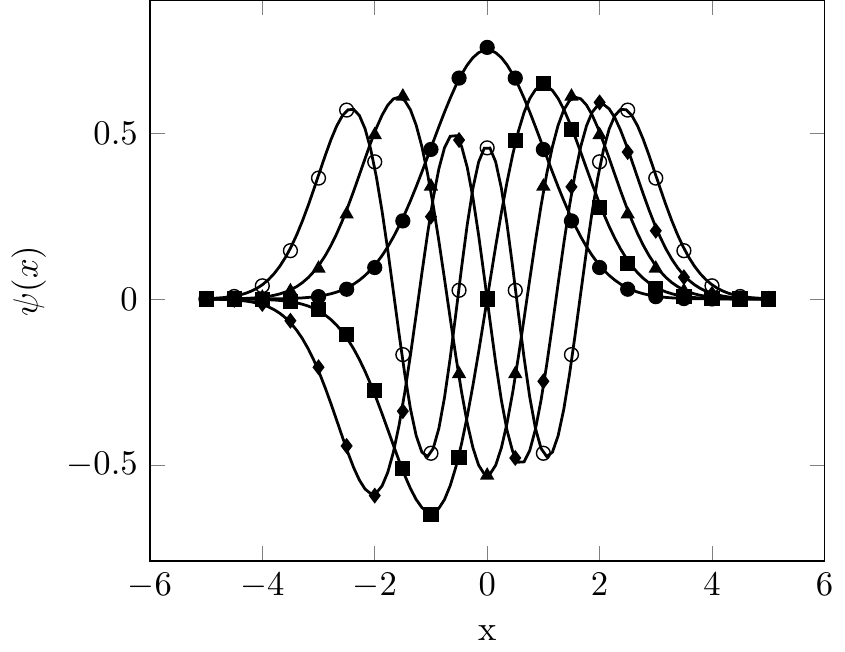}
\end{center}
\caption{Comparison of ground and excited state wavefunctions (symbols: state n = 1 ($\bullet$), n = 2 ($\blacksquare$), n = 3 ($\blacktriangle$), n = 4 ($\blacklozenge$), n = 5 ($\circ$)) with analytical wavefunctions (lines) for a particle in an harmonic oscillator potential.}
\label{fig-harmonic}
\end{figure}

\begin{figure}[p]
\begin{center}
\includegraphics[scale=1.0]{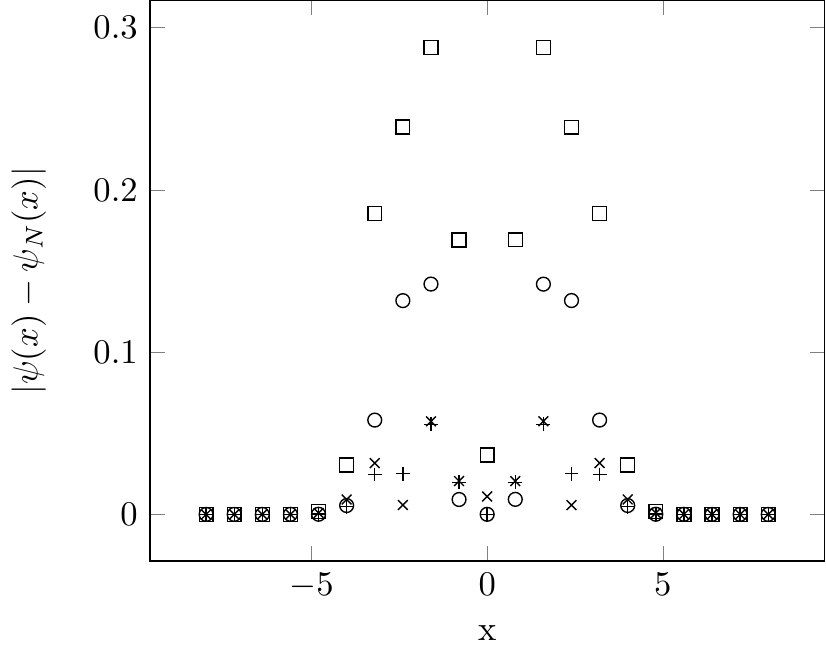}
\end{center}
\caption{Absolute errors of numerical wavefunctions $\psi_N(x)$ for state $n=3$ ($\circ$ symbols) and $n=4$ ($\square $) obtained by the FDTD method, and for state $n=3$ ($+$) and $n=4$ ($\times$) obtained by the NSFDTD method.}
\label{fig-error}
\end{figure}

Second application of the NSFDTD method is for a particle in an harmonic oscillator potential $V(x) = \frac{1}{2}\omega x^2$. The eigen energies are $E_n = (n+\frac{1}{2})\omega$. The wavefunctions are products of $\exp(-\omega x^2/2)$ and Hermite polynomials (details can be found in \cite{bransden2000, atkins2011}). Numerical results of eigen energies for an angular frequency $\omega = 1$, grid spacing $\Delta x = 0.8$ and $\Delta x = 0.5$ with computational length of 16 and 10 are given in Table \ref{tbl-harmonic}. Numerical wave functions for the first five states are shown in Fig.~\ref{fig-harmonic}. To compare accuracy of the FDTD and the NSFDTD results, absolute errors of numerical wavefunctions for states ($n=3$ and $n=4$) are shown in Fig.~\ref{fig-error}.     

In this application, the potential $V(x)$ and consequently the $W(x)$ are dependent on $x$. It is clearly shown, especially for excited states, the first iteration of the NSFDTD method is also found to improve significantly the standard FDTD method. Although the NSFDTD method is not exact in this case, its accuracy is still higher than the standard FDTD results. It is obviously shown that the numerical results become more accurate when the spacing is reduced. Because the wavefunction of the harmonic oscillator has exponential dependence, in order to get accurate results using the NSFDTD method we need to verify that the wavefunction has decayed to approximately zero near the computational boundary. This is to ensure that the boundary does not affect wavefunctions.

Comparison of numerical and analytical wavefunctions are shown in Fig.~\ref{fig-harmonic} and the sum of squared errors (SSE) of the wavefunctions are given in Table \ref{tbl-harmpsi}. The numerical wavefunctions are shown to be in agreement with the analytical wavefunctions. It is noted that the SSE of wavefunctions is decreasing as we increase the number of iteration. It is also shown that the wavefunctions are very small near the computational boundaries. Numerical wavefunctions obtained by the NSFDTD method is more accurate than by the FDTD method which is clearly shown by smaller absolute errors in Fig.~\ref{fig-error}.    

\begin{table*}[p]
\caption{\label{tbl-harmonic} Numerical eigen-energies for a particle in an harmonic oscillator potential computed by the FDTD method (standard scheme),  the NSFDTD after first (1st), second (2nd), third (3rd) and final iteration are compared with the exact values. The length of the computational domains are 16 and 10 for $\Delta x = 0.8$ and $\Delta x = 0.5$ respectively.}
\begin{center}
\begin{tabular}{@{}rrrrrrr}
\hline
No	&	FDTD	&	\multicolumn{4}{c}{NSFDTD}&	Exact	\\
	&		&	1st	&	2nd	&	3rd	&	final	&		\\
\hline
\multicolumn{7}{l}{Spatial Interval $\Delta x = 0.8$} \\
1	&	0.479077	&	0.497481	&	0.498004	&	0.498019	&	0.498019	&	0.5	\\
2	&	1.391838	&	1.483898	&	1.491834	&	1.492521	&	1.492586	&	1.5	\\
3	&	2.188108	&	2.435512	&	2.471583	&	2.476871	&	2.477781	&	2.5	\\
4	&	2.954963	&	3.357062	&	3.439530	&	3.456553	&	3.460975	&	3.5	\\
5	&	3.236360	&	4.050845	&	4.306467	&	4.384808	&	4.418670	&	4.5	\\
\hline
\multicolumn{7}{l}{Spatial Interval $\Delta x = 0.5$} \\
1	&	0.492062	&	0.499624	&	0.499705	&	0.499706	&	0.499706	&	0.5	\\
2	&	1.459805	&	1.497690	&	1.498910	&	1.498949	&	1.498951	&	1.5	\\
3	&	2.393936	&	2.491568	&	2.496825	&	2.497108	&	2.497125	&	2.5	\\
4	&	3.292528	&	3.479435	&	3.493647	&	3.494727	&	3.494816	&	3.5	\\
5	&	4.153118	&	4.459088	&	4.489421	&	4.492424	&	4.492753	&	4.5	\\
\hline
\end{tabular}
\end{center}
\end{table*}

\begin{table*}[p]
\begin{center}
\caption{\label{tbl-harmpsi} Same as Table \ref{tbl-harmonic} except that this is for the sum of squared errors of wave functions.}
\begin{tabular}{@{}rrrrrrr}
\hline
No	&	FDTD	&	\multicolumn{4}{c}{NSFDTD}\\
	&		&	1st	&	2nd	&	3rd	&	final\\
\hline
\multicolumn{6}{l}{Spatial Interval $\Delta x = 0.8$} \\
1 &   0.002640   & 0.001655   & 0.001631   & 0.001630   & 0.001630\\
2 &   0.015449   & 0.004473   & 0.003858   & 0.003807   & 0.003802\\
3 &   0.071309   & 0.011986   & 0.008113   & 0.007629   & 0.007548\\
4 &   0.134618   & 0.022206   & 0.011478   & 0.009849   & 0.009463\\
5 &   0.662104   & 0.118040   & 0.031514   & 0.017526   & 0.013284\\
\hline
\multicolumn{6}{l}{Spatial Interval $\Delta x = 0.5$} \\
1 &   0.000611   & 0.000381   & 0.000379   & 0.000379   & 0.000379\\
2 &   0.003254   & 0.000878   & 0.000827   & 0.000826   & 0.000826\\
3 &   0.011101   & 0.001747   & 0.001501   & 0.001489   & 0.001488\\
4 &   0.029720   & 0.002703   & 0.001972   & 0.001925   & 0.001921\\
5 &   0.068005   & 0.003694   & 0.002016   & 0.001898   & 0.001886\\
\hline
\end{tabular}
\end{center}
\end{table*}

\subsection{Morse Potential}

For application in atomic vibrations, the Morse potential well is usually chosen as the model potential. The Morse potential as an example for this paper is in the form of
\begin{equation}
V(x) = D[1-\exp(-(x-x_0))]^2
\end{equation}.
The analytical eigen energies are given by $E_n = \omega (n+\frac{1}{2})[1 - \omega (n+\frac{1}{2})/4D]$ with $\omega^2 = 2D$.

For numerical computations, the NSFDTD parameters are $x_0 = 2$, $\Delta x = 0.2$, the computational domain is $[0,8]$ and the Morse parameter $D = 50$ is used for simplicity. The resulting numerical eigenvalues are given in Table \ref{tbl-morse}. In this case, it is also found that the NSFDTD method is more accurate that the standard FDTD method.

\begin{table*}[p]
\caption{\label{tbl-morse} Numerical eigen-energies for a particle in a Morse potential computed by the FDTD method (standard scheme),  the NSFDTD after first (1st), second (2nd), third (3rd) and final iteration are compared with the exact values. The parameters used are $\Delta x = 0.2$, $x_0 = 2$ and the computational domain is at an interval [0,8].}
\begin{center}
\begin{tabular}{@{}rrrrrrr}
\hline
No	&	FDTD	&	\multicolumn{4}{c}{NSFDTD}&	Exact	\\
	&		&	1st	&	2nd	&	3rd	&	final	&		\\
\hline
1	&	4.754956	&	4.865394	&	4.867215	&	4.867245	&	4.867246	&	4.875	\\
2	&	13.365245	&	13.830904	&	13.851604	&	13.852525	&	13.852568	&	13.875	\\
3	&	20.758056	&	21.762245	&	21.828617	&	21.832999	&	21.833308	&	21.875	\\
4	&	27.087090	&	28.686514	&	28.818009	&	28.828801	&	28.829763	&	28.875	\\
5	&	32.484206	&	34.633288	&	34.831976	&	34.850315	&	34.852177	&	34.875	\\
\hline
\end{tabular}
\end{center}
\end{table*}

\section{Conclusion}
A numerical method known as the nonstandard (NS) FDTD method has been presented and validated using three potential wells. The numerical results of eigen energies and wavefunctions are compared with the standard FDTD and the analytical results. It has been shown that the NSFDTD method is an exact method for the case of a constant potential. The NSFDTD method has been also shown to perform better than the standard FDTD method for all cases.  

\section*{Acknowledgement}
This work is partially supported by Directorate Research and Community Service - Directorate General of Strengthening for Research and Development - Ministry of Research, Technology, and Higher Education  Republic of Indonesia.  


\end{document}